\pdfoutput=1
\documentclass[aps,prl,twocolumn,amsfonts,showpacs,superscriptaddress]{revtex4-1}
\usepackage{graphicx,color,transparent}
\usepackage{psfrag}
\usepackage{dcolumn}
\usepackage{amsmath}
\usepackage{amssymb}
\usepackage{latexsym}
\usepackage{times}
\usepackage{tikz}
\usepackage[plain]{fancyref}

\definecolor{mypurple}{RGB}{153,61,113}
\definecolor{myblue}{RGB}{63,61,153}
\definecolor{myokker}{RGB}{153,140,61}
\definecolor{mygreen}{RGB}{61,153,86}
\definecolor{mymarine}{RGB}{61,90,153}
\definecolor{mycyan}{RGB}{0,255,255}

\begin{document}

%%%%%%%%%%%%%%%%%%%%%%%%%%%%%%%%%%%%%%%%%%%%%%%%%%%%%%%%%%%%
%%%%%%%%%%%%%%%%%%%%%%%%%%%%%%%%%%%%%%%%%%%%%%%%%%%%%%%%%%%%
%\newcommand{\rem}[1]{{\color{red} \textbf{REMARK: #1}}}

\def\bea{\begin{eqnarray}}
\def\eea{\end{eqnarray}}
\def\beq{\begin{equation}}
\def\eeq{\end{equation}}
\def\f{\frac}
\def\k{\kappa}
\def\e{\epsilon}
\def\ve{\varepsilon}
\def\be{\beta}
\def\D{\Delta}
\def\h{\theta}
\def\t{\tau}
\def\a{\alpha}

\def\rk{\rho^{ (k) }}
\def\rek{\rho^{ (1) }}
\def\cek{C^{ (1) }}
\def\rz{\bar\rho}
\def\rt{\rho^{ (2) }}
\def\rtb{\bar \rho^{ (2) }}
\def\trk{\tilde\rho^{ (k) }}
\def\trek{\tilde\rho^{ (1) }}
\def\trz{\tilde\rho^{ (0) }}
\def\trt{\tilde\rho^{ (2) }}
\def\r{\rho}
\def\tD{\tilde {D}}

\def\s{\sigma}
\def\kb{k_B}
\def\la{\langle}
\def\ra{\rangle}
\def\nn{\nonumber}
\def\up{\uparrow}
\def\dn{\downarrow}
\def\S{\Sigma}
\def\dg{\dagger}
\def\d{\delta}
\def\p{\partial}
\def\l{\lambda}
\def\L{\Lambda}
\def\G{\Gamma}
\def\o{\Omega}
\def\w{\omega}
\def\g{\gamma}

\def\cLf{{\cal L}_1}
\def\cLo{{\cal L}_0}

\def\noi{\noindent}
\def\a{\alpha}
\def\d{\delta}
\def\p{\partial} 

\def\la{\langle}
\def\ra{\rangle}
\def\e{\epsilon}
\def\n{\eta}
\def\g{\gamma}
\def\break#1{\pagebreak \vspace*{#1}}
\def\hf{\frac{1}{2}}

\title{Pumping single-file colloids: Absence of current reversal}
\author{Debasish Chaudhuri} 
\affiliation{Indian Institute of
  Technology Hyderabad, Yeddumailaram 502205, Andhra Pradesh, India }
\email{debc@iith.ac.in}

\author{Archishman Raju} 
\affiliation{ Department of Physics, Cornell University, Ithaca, 
 New York 14853, USA}
%\email{debc@iith.ac.in}

\author{Abhishek Dhar} 
\affiliation{International Centre for Theoretical Sciences, TIFR, Bangalore - 560012}
\email{abhishek.dhar@icts.res.in}

\date{\today}

\begin{abstract}
We consider the single-file motion of colloidal particles  interacting via short-ranged repulsion and placed in  a traveling wave potential, that varies periodically in time and space. Under suitable driving conditions, a directed time-averaged flow of colloids is generated. We obtain analytic results for the model 
using a perturbative approach to solve the Fokker-Planck equations. The predictions show good agreement with numerical simulations. 
We find peaks in the time-averaged directed current  as a function of driving 
frequency, wavelength and particle density and discuss possible experimental realizations. Surprisingly, unlike a closely related exclusion dynamics on a lattice, the directed current in the present model does not show current reversal with density. A linear response formula relating current response to equilibrium correlations is also proposed.
\end{abstract}

%{Nonequilibrium and irreversible thermodynamics}
%{Fluctuation phenomena, random processes, noise, and Brownian motion} 
%{Transport processes}

\maketitle
In single-file motion, colloidal particles are confined to move in a narrow channel such that they cannot overtake each other.  
This  was first studied by Hodgkin and Keynes~\cite{hodgkin1955} while trying to describe ion transport in biological
channels. 
One of  the most interesting features of single-file motion is the sub-diffusive behavior that individual particles exhibit, and has been extensively studied both theoretically~\cite{rodenbeck1998,lizana2008,barkai2009,roy2013}  and experimentally~\cite{Hahn1996,Kukla1996,Wei2000,Lutz2004,lin2005,das2010}. 
An exciting question  has been that of obtaining directed particle currents in such single-file systems in closed geometries, for example colloidal particles moving in a circular micro-channel.  
Using  periodic forces that vanish on the average, it has been possible to drive  particle currents in a 
unidirectional manner. These are referred to as Brownian ratchets and may, for example, be achieved  through continual switching on and off of a spatially asymmetric potential profile~\cite{Julicher1997,Reimann2002}.
Such phenomena have been studied experimentally using suitably constructed electrical
gating~\cite{Rousselet1994,Leibler1994,Marquet2002}, and with the help of laser tweezers~\cite{Faucheux1995,Faucheux1995a,Lopez2008}.  
Intracellular motor proteins like kinesin, myosin that move on respective filamentous tracks~\cite{Reimann2002},
or  ${\rm Na}^+$-, ${\rm K}^+$-ATPase pumps associated with the cell-membranes~\cite{Gadsby2009}, %have been proposed as 
are examples of naturally occurring stochastic pumps.   
With a few exceptions~\cite{Derenyi1995,Derenyi1996,Aghababaie1999,Slanina2008a,Slanina2009,savel2004},  
most theoretical studies of Brownian ratchets  focused  on systems of non-interacting particles. 

Recently a  model of  classical stochastic pump~\cite{Chaudhuri2011, Marathe2008, Jain2007} has been proposed,  similar to those used in the study of quantum pumps~\cite{Brouwer1998,  Citro2003}. %Instead of asymmetric potentials that are switched on and off either periodically or stochastically, 
Unlike Brownian ratchets, in these pump models, the colloidal particles are driven by a traveling wave potential. Thus, while typical ratchet models consider particles in a potential of the form $V(x,t)=f(x)g(t)$, the pump model considers a form such as $V(x,t)=V_0 \cos(q x -\omega t)$.
In Ref.~\cite{Chaudhuri2011}, the dynamics of  colloidal particles with short ranged repulsive interactions, and  confined to 
move on a ring in the presence of an external space-time varying potential,
was studied by considering a discretized version. In the discrete space model, particles moved on a lattice with the  exclusion constraint that sites cannot have more than one particle and hopping rates between neighboring sites depended on the instantaneous potentials on the sites. This roughly mimics the over-damped Langevin dynamics of hard-core particles that is expected to be followed  by 
sterically stabilized colloids.  As expected, the traveling wave potential resulted in a DC particle current in the ring. An intriguing result 
was that the system  showed a  current direction-reversal on increasing the  density beyond half-filling. This behavior was an outcome of the particle-hole
symmetry of the discrete model~\cite{Chaudhuri2011}. 
Current reversal has been observed in  subsequent theoretical studies~\cite{Dierl2014,pradhan14}.
Further interesting properties of this model, including a detailed phase diagram, were recently obtained for the case where the system was connected to reservoirs and a biasing field applied~\cite{Dierl2014}. 
{General conditions for pumping to occur have recently been discussed in \cite{Rahav2008, Mandal2011,Asban2014}.}

An important question is as to how much of the interesting qualitative features, seen in the lattice model, remain valid for real interacting colloidal particles executing single-file Brownian dynamics. This is one of the main motivations of this Letter. Here we consider the effect of a traveling wave potential on such particles which can be described by Langevin dynamics. 
\begin{figure}[t]%[htbp]
\begin{center}
\includegraphics[width=7 cm]{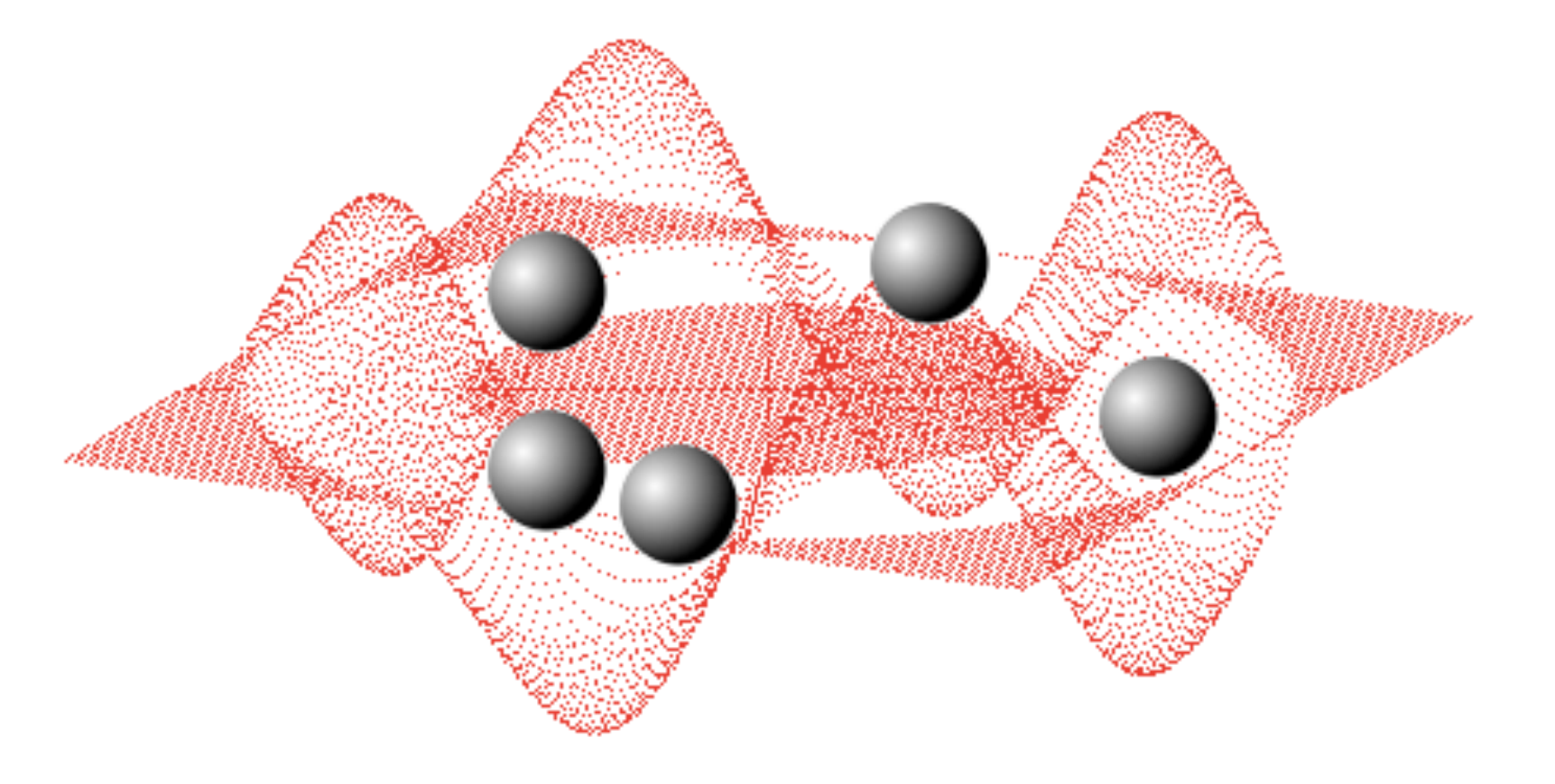} %{graph1.eps} %{Wj_free_pump.eps}  
\caption{(Color online) 
A circular potential trap, which confines the motion of colloids in one dimension, is denoted by the white annulus. The colloidal particles are shown by dark spheres. 
The oscillatory profile indicates a time-frozen version of the traveling wave potential $V = V_0 \cos(\w t - q x)$.}
\label{fig:cartoon}
\end{center}
\end{figure}
Numerical and some analytic results based on the solution of the Fokker-Planck equation are presented. We derive a linear response formula for 
the DC current in terms of equilibrium correlation functions. 
We find that, unlike the lattice version~\cite{Chaudhuri2011,Marathe2008,Jain2007}, there is no current-reversal in this  
system. A proposal for possible experimental realization of particle pumping in colloidal systems, using traveling waves, is discussed.

%model
We consider $N$ colloidal particles that are confined to move on a one-dimensional ring of length $L$. The particles interact via potentials $U(x)$ 
 that are sufficiently short ranged that we  take them to be only  
between nearest neighbors. In addition, a weak traveling wave potential of the form $ V(x,t) = \l k_B T \cos(\w t-q x)$ with $\l < 1$, 
acts on each particle. Let $x_i$, $i=1,2,\ldots,N$ denote the 
positions of the particles along the channel. 
Then the over-damped Langevin equations of motion of the system are given by
\begin{align}
 \f{dx_i}{dt}&= -\mu \f{\partial \cal U}{\partial x_i}+ \eta_i,
\label{lange} \\
{\rm where} ~{\cal U}&=\sum_{i=1}^N V(x_i) +  \sum_{i=1}^N U(|x_i - x_{i+1}|) \nn 
\end{align}
is the total potential energy of the system and 
$\eta_i(t)$ is white Gaussian noise with 
$\la \eta_i \ra=0$,~$\la \eta_i(t) \eta_j(t') \ra = 2 D \d_{i,j}\d(t-t'),~ D=\mu \kb T$ is the diffusion constant, 
$\mu$ the mobility, $\kb$ the Boltzmann constant and $T$ the ambient temperature. We have taken periodic boundary conditions $x_{N+1}=L+x_1$.
 %AD Check

 Denoting the joint probability distribution of the $N$-particle system as $P({\bf x},t)$ with 
 ${\bf x}=(x_1, x_2,\dots,x_N)$, the Fokker-Planck equation governing its time
 evolution is 
 \bea
 \p_t P = \sum_i \p_{x_i} [ D \p_{x_i} P + \mu P \p_{x_i} {\cal U}]~.
 \label{eq:fp}
 \eea
 The  one-point distribution for the $i^{\rm th}$ particle is given by $P^{(1)}_i(x_i,t) = \int d x_1 d x_2 \dots d x_{i-1} dx_{i+1}\ldots d x_N P({\bf x},t)$. 
Similarly let $P^{(2)}_{i,i+1} (x_i,x_{i+1})$ be the two-point distribution obtained from $P({\bf x})$ by integrating out all  coordinates other than $x_i,x_{i+1}$. 
Let us then define the averaged distributions 
$P^{(1)}(x,t) = \f{1}{N} \sum_i P^{(1)}_i(x,t)$ and  
$P^{(2)}(x,x',t) = \f{1}{N} \sum_i P^{(2)}_{i,i+1}(x,x',t) $.
 Integrating the $N$-particle
 Fokker-Planck equation one finds a BBGKY hierarchy of equations, the first of which is
 \begin{align}
 \p_t P^{(1)}(x,t) =& -  \p_x J~, \label{intPeq} \\
 {\rm where}\, J  =& - D \p_x  P^{(1)}(x,t)  -\mu  \p_x V P^{(1)}(x,t) \nn \\
    & - \mu \int d x'  \p_x U(|x-x'|) P^{(2)}(x,x',t) . \nn
  \end{align}
The  local density of particles is given by $ \rho(x,t) = N P^{(1)}(x,t)$, and  
the corresponding current density is $j(x,t)=N J(x,t)$.
The time and space averaged directed current in the system is given by
\bea
j_{\rm DC} = \f{1}{\t L} \int_0^{\t} dt \int_0^L dx j(x,t)~,
\eea
where $\tau=2 \pi /\omega$.

\begin{figure}[t]%[htbp]
\begin{center}
\includegraphics[width=8.6 cm]{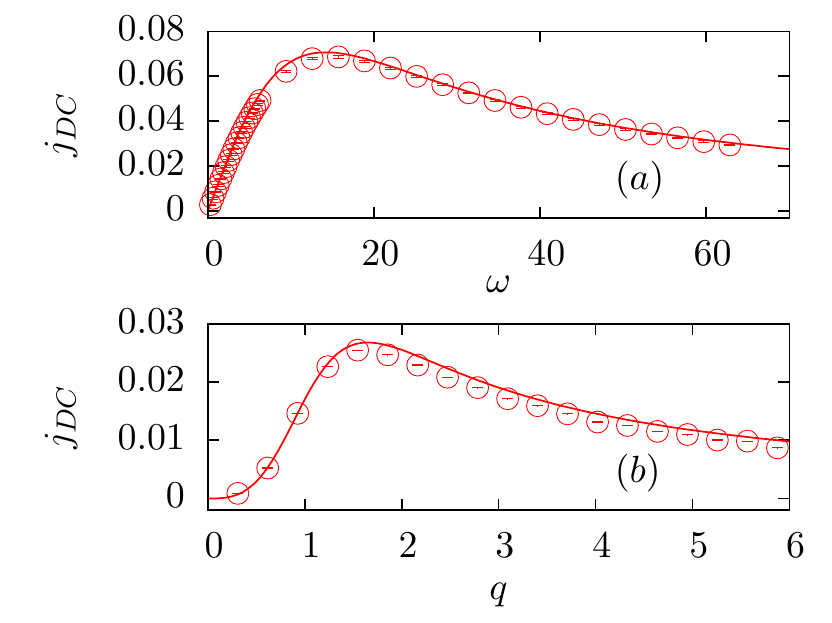}
\caption{(Color online) 
Directed current $j_{\rm DC}$ as  a function of ($a$) driving frequency $\w$ and ($b$) driving wave-number $q$ in the non-interacting system.  
The points denote Langevin dynamics simulation results of free particles and the solid lines  are the analytic prediction of Eq.~(\ref{jdc_free}). The parameters used are: number of particles $N=128$,  
mean density $\r_0=0.3$,  potential strength $\l=0.5$, diffusion constant $D=1$ and temperature $k_BT=1$. In (a)  $q=1.2\,\pi$ while in (b) $\omega=\pi/2$.}
\label{nonint}
\end{center}
\end{figure}

{\em Non-interacting system:}
We first analyze the non-interacting system ($U=0$). %for which we analyze the model  using a perturbative expansion of $\r(x,t)$  in  $\l$. 
The Fokker-Planck equation for $\r(x,t)$ is 
\bea
\p_t \r(x,t) + \p_x j =0,~ ~ j(x,t) = -D \left[ \be V' + \p_x  \right] \r(x,t)
~\label{eq:jxt}
\eea
with $V'=\p_x V$. 
%We expand the density  as
We expand the density in a perturbative series in small parameter  $\l$ as
\beq
\r(x,t) = \r_0 + \sum_{k=1,2,\dots} \l^k \r^{(k)} (x,t),
\eeq
where $\r_0 = N/L$ is the  mean density of particles.
The mean directed current $j_{\rm DC}$ gets a contribution only from the drift part of the current in Eq.~(\ref{eq:jxt}), which to leading order is given by  $-D \be V' \lambda \r^{(1)}$. The time evolution for $\r^{(1)}$ is given by
\bea
\p_t \r^{(1)}  - D \p_x^2 \r^{(1)}  &=&  \r_0 D \p_x^2 (\be V/\lambda)~, 
\eea
and this has the time-periodic steady state solution 
\bea
\r^{(1)} &=& \r_0 q^2 D~ {\rm Re} \left[ \f{e^{i(q x -\w t)}}{i\w - D q^2} \right]~. 
\eea
Thus, to leading order in the perturbation series in $\l$, the time averaged directed current is 
\bea
j_{\rm DC} &=& \f{1}{\t L}\int_0^{\t} dt \int_0^L dx (-\be D \l   \r^{(1)}  \p_x V) \nn\\
&=& \f{\l^2 \r_0}{2} \f{D^2 q^3 \w}{D^2 q^4 + \w^2} .
\label{jdc_free}
\eea
As expected, the current has a linear dependence on particle density $\rho_0$.
The dependence on driving frequency $\omega$ and wave-number $q$ are plotted in
 Fig.~(\ref{nonint}) where we also show a comparison of the results from the analytic perturbative theory with those from direct numerical simulations for $\lambda=0.5$. We see that there is excellent agreement even for this, not very small,  value of $\lambda$.

\begin{figure}[t] % [htbp]
\begin{center}
\includegraphics[width=8.6 cm]{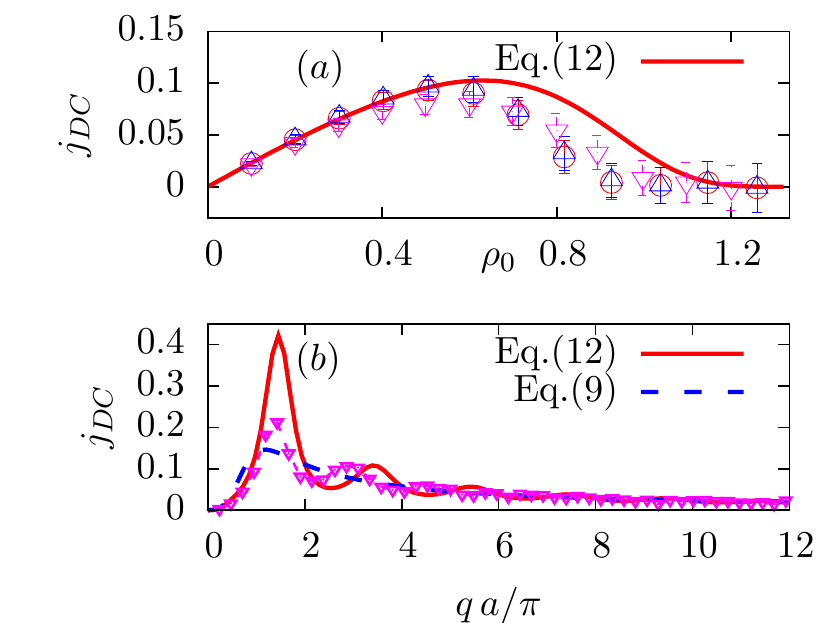}  %{graph33.pdf}
%{\input{graph33}} %{fig3.pdf_tex}%{graph3}
\caption{(Color online) 
Directed current $j_{\rm DC}$ as  a function of ($a$)~mean density $\r_0$, and ($b$)~wave number $q$ in the interacting system. The points denote Langevin dynamics simulation results of particles interacting via WCA ($\circ$), soft-core ($\triangle$), and Fermi-function step potentials ($\triangledown$). 
%and the lines are  guide to eye. 
The solid lines in ($a$) and ($b$) are plots of Eq.~(\ref{jdc_mf}) with $a=0.75$. %$a=0.64$. % and $\a=\kb T[\r_0 a/(1-\r_0 a)]$. % and $\alpha= 160$. %26.92$. %$a=0.1\s$. %$u_0=0.47\e\s$. 
Parameters are: $\w=14$, and $\l=0.5$; in ($a$)~$q=1.2\pi$, in ($b$)~$\r_0 = 0.55$. 
%All other parameters are the same as in Fig.~\ref{fig:jdcw}. 
}
\label{fig:jdcro}
\end{center}
\end{figure}

%{\em Perturbative calculation of the current: interacting system}
{\em Interacting system:}
Let us consider  a hard-core interaction between the particles  defined through the potential $U(x) = \infty$ if $|x| < a$ and $0$ otherwise.
Since $N P^{(1)}=\rho(x,t)$ gives the density of particles and defining the pair
distribution function $g(x,x',t)$ through the relation $N P^{(2)}=\rho(x,t) g(x,x',t)$, we see that Eq.~(\ref{intPeq}) can equivalently be written as
 \begin{align}
 \p_t &\rho(x,t) = D \p^2_x \rho + \mu \p_x \left[ V(x,t) \rho(x,t)\right] \nn \\
    & + \mu \p_x \left[ \rho(x,t)  \int d x'  \p_x U(|x-x'|) g(x,x',t)\right]~ . \label{intreq}
  \end{align}
On expanding $\rho(x,t)$ and $g(x,x',t)$ as perturbation series in $\lambda$ we find that the resulting equations {\emph{do not close}} at successive orders. This is different from the case of the discrete systems studied in \cite{Chaudhuri2011,Marathe2008,Jain2007} where the perturbative solution works even in the presence of interactions. We  thus need to make further approximations before applying the perturbation theory. It turns out that a mean-field description of the interaction term in Eq.~(\ref{intreq}) makes the problem tractable. 
%%%%%%%%
The pair correlation function $g(x,x',t)$ gives the probability of finding a particle at $x'$ given that there is a particle at $x$, while $-\p_x U(|x-x'|)$ is the force on the particle at $x$ due to a particle at $x'$. Hence the integral  ${\cal I}=\int d x'  [- \p_x U(|x-x'|)] g(x,x',t)$ has the interpretation of being the average force on a particle located at $x$. Next we note that for a hard rod centered at $x$, the force is localized at the points $x\pm a $, hence we can approximate the average force by the pressure difference between these points, i.e., {${\cal I} = \Pi (x-a)-\Pi (x+a)$}.
%the integral ${\cal I}$ can be replaced by   $\Pi (x-a)-\Pi (x+a)$. 
Here we assume that $\Pi(x,t)$ is the instantaneous local equilibrium pressure, 
finally we relate this pressure to the density $\rho(x,t)$ through the equilibrium relation  $\Pi=k_B T \rho/[1-\rho a]$~\cite{Chowdhury2000}. 
Using this form of the interaction term and expanding $\rho(x,t)$ to first order in $\l$, the time evolution equation for $\rho^{(1)}$ is
\begin{align}
\p_t \r^{(1)}  &- D \p_x^2 \r^{(1)}  =  \r_0 D \p_x^2 (\be V/\lambda)  \nn \\
&+  D \alpha \p_x \left[ \rho^{(1)}(x+a,t) - \rho^{(1)}(x-a,t) \right]~, \nn  
\end{align}
where $\alpha=  \rho_0/(1-\rho_0 a)^2$.
%
%As before, 
The time-periodic steady state solution of this equation is given by
\bea
\r^{(1)} = \r_0 q^2 D~ {\rm Re} \left[ \f{e^{i(q x -\w t)}}{i\w - D q^2 - 2 D \a  q \sin(q a)} \right]~.
\eea
This leads to, up to order $\l^2$ in perturbation series, the following  average current
\begin{equation}
j_{\rm DC} = \f{\l^2 \r_0}{2} \f{D^2 q^3 \w}{D^2 [q^2+2 \a  q \sin(q a)]^2 + \w^2}~.
\label{jdc_mf}
\end{equation}
This is the first main result of our paper.
We now see a non-trivial dependence on particle density $\rho_0$ and wave-number $q$.  For a fixed density there is enhancement of particle current at some $q$ values [see Fig.~(\ref{fig:jdcro})]. The current vanishes at the full-packing density $\rho_0=1/a$, as expected. However, unlike the lattice model, {\emph {there is now no current reversal}}.   
In the discrete lattice model of symmetric exclusion process driven by a potential $\l\cos(\w t-\phi n)$ with $n$ a lattice site and $\phi=q a$~\cite{Chaudhuri2011}  , 
 \bea
j_{\rm DC} %&=& \f{\l^2 f_0 \w}{2} (q_0 -2 k_0) {\rm Im} \left[\f{\sin \phi}{i\w - f_0 \e_\phi}\right] \crcr
&=& \l^2  f_0^2  (q_0 -2 k_0)  \f{\w  \sin\phi(1-\cos\phi)}{\w^2+4 f_0^2(1-\cos\phi)^2},
\label{eq:excl}
\eea
where $f_0=D/a^2$, and $(q_0 -2 k_0) =\eta(1-\eta)(1-2\eta)$ in the large $L/a$ limit with $\eta=\rho_0 a$ the packing fraction. 
The dynamics had particle-hole symmetry leading to current reversal at $\eta=1/2$. 
{In the continuum dynamics performed by colloidal particles, there is no such particle- hole symmetry. Note that} 
%and thus the current does not show reversal of direction with increasing density, as captured by Eq.(\ref{jdc_mf}).}
the continuum limit of Eq.(\ref{eq:excl}) with $a/L \to 0$, $\phi = q a \ll 1$ and $\eta=\r_0 a \ll 1$ leads to
the result for non-interacting %system 
{colloids} Eq.~(\ref{jdc_free}). {Presumably, the correct discrete model that one needs to consider, in order to get the correct continuum limit, is one  where particles occupy a finite number (large) of sites  and then one has to take appropriate limts.}

{\em Langevin dynamics simulations:}
To test our analytic predictions, we performed Langevin dynamics simulations of the model using Euler integration of Eq.~(\ref{lange}). 
The time scale is set by $\t_D = a^2/D$. For the non-interacting system, we used an integration time-step $\d t = 10^{-2} \t_D$. 
For the interacting single-file case, in order to avoid unphysical particle crossings at large densities,  we used $\d t = 10^{-5}\t_D$. 
A total of $N=128$ particles were simulated. The particle flux is averaged over the system, and 
%measurements across several cross-sectional lines within the system and 
over a time period $100 \tau$ where $\tau=2\pi/\w$. 
The particle current is further averaged over $100$ realizations. The fluctuations over realizations provide the errors in the measured currents. 
%To model the hard-core particle interactions and  
To check the robustness of our results, we considered a number of smooth potentials to model the short-ranged inter-particle repulsion: (a) Weeks-Chandler-Anderson (WCA) potential~\cite{Weeks1971}  $\be U(x) = 4 [ (\s/x)^{12} - (\s/x)^6 + 1/4]$ if $|x|<2^{1/6}\s$ else $U=0$, (b) Soft core potential $\be U(x) = (\s/x)^{12} - 2^{-12}$ if $|x|<2\s$ else $U=0$,  and (c)~Fermi-function step potential $\be U(x) = A/[\exp((x-a)/w)+1]$ with $A=100$, $w=0.02 \s$ and $a=0.75 \s$. 
In the simulations $\kb T = 1/\be$ and $\s$ set the energy and length scales respectively. 
The simulation data for all the three potentials agree with each other within numerical errors (Fig.~\ref{fig:jdcro}($a$)). 
They show a non-monotonic variation with density, with maximal current near $\r_0\s=0.55$. A plot of  Eq.~(\ref{jdc_mf}) with $a=0.75\s$ shows
qualitative agreement with numerical data. Fig.\ref{fig:jdcro}($b$) shows $j_{\rm DC}$ as a function of driving wave-number $q$ in a system of particles interacting via Fermi-function step potential. 
Multiple maxima in $j_{\rm DC}$ appears, in qualitative agreement with Eq.~(\ref{jdc_mf}). A comparison with Eq.~(\ref{jdc_free}) shows another intriguing feature, directed current
in presence of repulsive interaction can be higher than that of free particles.    
%
%
%\begin{figure}[t] % [htbp]
%\begin{center}
%{\input{graph55}} %{fig3.pdf_tex}%{graph3}
%\caption{(Color online) 
%The headway distribution between consecutive particles at various densities $\r_0$ between $0.7$ and $1.2$. The distributions are sharply peaked at high densities $\r_0 \gtrsim 1$.  For small densities the 
%distributions have exponential tail, that gets longer with reducing densities.
%}
%\label{fig:head}
%\end{center}
%\end{figure}
%

{\emph{ Linear response theory}}: Even though the current response is ${\cal O}(\lambda^2)$ and hence nonlinear in the perturbation, we note that it was obtained from 
the first order change in the density and hence should be calculable from linear response theory. We now show that it is indeed possible to express the current response to the perturbing traveling wave potential, in terms of equilibrium 
correlation functions of various forces,  using linear response theory.
Let us write the equation of motion in the form $\dot{x}_i=\mu[F_i(t)+ f_i]+\eta_i$, where $F_i(t)=-\p_{x_i}V(x_i,t)$ and $f_i=-\p_{x_i}U(|x_i-x_{i+1}|)-\p_{x_i} U(|x_i-x_{i-1}|)$ is the total force on $i^{\rm th}$ particle from its neighbors. We see that 
the total current is given by $\int_0^L dx \la j(x) \ra = \sum_{i=1}^N \la \dot{x}_i \ra = \mu \sum_{i=1}^N \la F_i \ra$, 
where $\la F_i \ra = \int d {\bf x}  F_i(x_i,t) P({\bf x},t)$~. The long time solution $P({\bf x},t)$  can be obtained from
perturbation theory. The Fokker-Planck equation for $P$, given by Eq.~(\ref{eq:fp}), can be expressed as $\p_t P = \cLo P + \cLf P$ where  $\cLo = \sum_i [D \p_{x_i}^2 - \mu \p_{x_i} f_i ]$ 
 and the external perturbation is $\cLf=-\sum_i \mu \p_{x_i} F_i$. Writing $P = P_0 + P_1$, where $P_0 = \exp[-\be \sum_i U(x_i, x_{i+1})]/Z$ is the equilibrium state, one gets to $\cal O(\l)$, 
 $P_1({\bf x},t) = \int_{-\infty}^t d t' e^{\cLo (t-t')} \cLf P_0({\bf x})$~.
 Using this, to leading order in $\l$, one obtains 
\begin{align}
&\la F_i \ra = -\mu \int_{-\infty}^t dt' \int d{\bf x} F_i(t) e^{\cLo (t-t')} \sum_j  \p_{x_j} [F_j(t') P_0({\bf x})]  \nn \\
&= -\mu \int_{0}^\infty du  \big\la F_i(t) e^{\cLo u} \sum_j \left[\p_{x_j} F_j(t-u) +\beta F_j(t-u) f_j\right] \big\ra_{0}, \nn
\end{align}
where $\la\ldots\ra_0$ refers to an equilibrium average, and  the time-dependence in $F_i(t)=F_i(x_i,t)$ only refers to the explicit time-dependence of the external force. 
Using the fact that $F_i=-\lambda k_B T q \sin (q x_i - \omega t)$ and  that $\la A(t) B(0) \ra= \int d {\bf x} A({\bf x}) e^{\cLo t} B({\bf x}) P_0$ we get 
\begin{align}
&\la F_i \ra =-\l^2 q^2 \mu (k_B T)^2 \int_0^\infty du \big\la \sin (q x_i(u)-\omega t) \nn \\ 
&  \times \sum_j \left[q \cos (q x_j -\omega (t-u)) + \beta f_j \sin (q x_j -\omega (t-u) ) \right] \big\ra_0~.  \nn 
\end{align}
Finally, after averaging over a time period we get for the DC current:
\begin{align}
j_{DC}&=\sum_{i,j} \f{-(\l q \mu k_B T)^2}{2 L} \int_0^\infty dt \Big[ q \big\la   \sin [q (x_i(t)-x_j)-\omega t ] \big\ra_0 \nn \\
&  + \big\la \beta f_j \cos [q (x_i(t)-x_j)-\omega t] \big\ra_0 \Big]~. \label{LR}
\end{align}
This linear response formula, relating the DC current to equilibrium correlation functions, is  the second main result of this paper.

{\em Possible experiment:}
Using oscillating mirrors, it is possible to move a strongly focused infrared laser beam along a circle to constrain $\s \approx 1 \mu$m sized polystyrene 
spheres to move along a circle~\cite{Faucheux1995}. The steric interaction between polystyrene beads would lead to single-file motion. Using similar techniques as in \cite{Faucheux1995}, one can generate a  cosine  potential by passing the laser through an appropriately graded filter. Finally, a traveling wave potential can be formed by  rotating the filter at the required frequency. 
%At room temperature let us take the diffusion coefficient $D  \approx 1 \mu{\rm m}^2{\rm s}^{-1}$. 
If we choose the driving force wavelength to be few particle sizes so that $qa \approx 1$ then the optimal driving frequency is $\omega \approx D q^2 \approx D/a^2 \approx 1 Hz$,
using $D  \approx 1 \mu{\rm m}^2{\rm s}^{-1}$ at room temperature. This leads to a current $j_{\rm DC}\approx 0.05 {\rm s}^{-1}$ at a density $\rho_0 a \approx 0.5$. This is comparable to the currents obtained using the flashing ratchet mechanism in \cite{Faucheux1995}.

In summary, we investigated the dynamics of interacting colloidal particles confined to move in a narrow circular channel and driven by a traveling wave potential. 
Using a combination of mean-field type assumptions and perturbation theory, analytic results were obtained for the average particle current in the channel. This compares quite well with simulation results. We have also proposed a linear response formula relating the current response to equilibrium correlations. This relation opens up further analytic possibilities.  
The current shows peaks  as a function of driving frequency and wave number, and also the particle density. The  current vanishes as we approach the close packing limit and, rather surprisingly, does not show current reversal unlike what is seen in  studies of discrete versions of this model~\cite{Chaudhuri2011}. From 
the point of view of experiments, the pumping of colloidal particles in narrow channels using traveling wave potentials looks very accessible and could have potential applications.   

%
%\section{Outlook}
%In order to create the external periodic driving, one may use laser traps, or oscillatory pressure, or electro-kinetic coupling.
%
%Let us consider one dimensional motion of sterically stabilized colloids of diameter $\s \approx 1 \mu $m moving
%in water of viscosity $\nu \approx  10^{-3}$ ${\rm N s/m}^2$ that corresponds to Stokes-Einstein 
%self-diffusion constant $D = \kb T/3\pi\nu \s \approx 0.4\, \mu {\rm m}^2/$s, and thus $\t \approx 2.5\,$s.
%For a colloidal density $0.3\,\mu {\rm m}^{-1}$, at the 
%driving phase factor $q=1.2 \pi$ the resonance in directed current
%would be obtained at a driving frequency $\w/2\pi \approx 1\,$Hz.
%The corresponding maximal current will give rise to
%a mean colloidal velocity $\s j_{\rm DC} \approx 0.03 \l^2 \, \mu {\rm m\,s}^{-1}$, which with a potential strength $0.5\,\kb T$ 
%leads to $0.007\,\mu {\rm m\,s}^{-1}$.

\acknowledgements
DC and AR thank RRI Bangalore for hospitality where this work was initiated. DC thanks MPI-PKS Dresden for hosting him at various stages
of this work, and ICTS-TIFR Bangalore for hospitality while writing the paper.

\bibliographystyle{prsty}
%\bibliography{pump}

\end{document}